\documentclass[10pt,letterpaper,twocolumn]{article} 

\usepackage{ol2}
\usepackage[draft]{hyperref}
\usepackage{amsmath}

\begin{document}

\twocolumn[ 

\title{Light wheel buildup using a backward surface mode}

\author{R\'emi Poll\`es, Antoine Moreau, and G\'erard Granet}
\address{Clermont Universit\'e, Universit\'e Blaise Pascal, LASMEA, BP 10448, F-63000 Clermont-Ferrand}
\address{CNRS, UMR 6602, LASMEA, F-63177 Aubi\`ere}

\begin{abstract}
When a guided mode is excited in a dielectric slab coupled to a backward surface wave at the interface between a dielectric and a left-handed medium, light is confined in the structure : this is a light wheel. Complex plane analysis of the dispersion relation and coupled-mode formalism
give a deep insight into the physics of this phenomenon (lateral confinement and the presence of a dark zone).
\end{abstract}


]

Recently, a lamellar structure consisting of a conventional dielectric layer coupled to a left-handed material (LHM) layer has been proposed to confine light. Because of contradirectional power flows in the two layers, an exotic localized mode called "light wheel" can be excited \cite{tichit, ye}. In this letter, we present our study of a new type of light wheel that uses a backward surface mode at the interface between an LHM and a right-handed material (RHM). The dispersion relation of the structure and the coupled-mode formalism allows us to describe and explain the field distribution in the structure.

Let us first consider a surface wave propagating along an interface between two semi-infinite media: an LHM and a RHM. The LHM is caracterized by its dielectric permittivity $\varepsilon_3$ and its magnetic permeability $\mu_3$, which are both negative. The permittivity of the RHM is $\varepsilon_1$, its permeability is $\mu_1$. Because of unusual properties of LHM, such an interface supports surface guided modes which can be backward (i.e., present opposite phase and group velocities) \cite{darma, shadri}.
In TM polarization, the dispersion relation for surface waves propagating along this interface can be written as \cite{shadri}
\begin{equation}
\alpha^2 = k_0^2\ \mu_1\ \varepsilon_1\ \frac{X\left(X-Y\right)}{\left(X^2-1\right)},
\label{eq:plasmon}
\end{equation}
where $X=\frac{|\varepsilon_3|}{\varepsilon_1}$, $Y=\frac{|\mu_3|}{\mu_1}$, $\alpha$ is the propagation constant and $k_0$ is the wavenumber in the vacuum. A backward surface wave is supported provided the inequality 
\begin{equation}
1<X<1/Y
\label{eq:condition}
\end{equation}
is verified.

The light wheel develops from the contradirectional coupling between this backward mode and a forward guided mode. This coupling appears when the dispersion curves of the two waveguides cross each other (i.e. when they are under phase matching conditions). Even when taking the dispersive character of the LHM into account, such a coupling is thus not difficult to obtain as long as the condition (\ref{eq:condition}) is fulfilled. At a given wavelength $\lambda$, let us consider, for instance, that $\varepsilon_3=-0.8$ and $\mu_3=-1.5$. Because that satisfies the previous conditions, a backward surface wave between such an LHM and the air ($\varepsilon_1=1$, $\mu_1=1$) exists for $\alpha=\alpha_0=1.2472k_0$ according to Eq. (\ref{eq:plasmon}).
 
\begin{figure}[htb]
\centerline{\includegraphics[width=8cm]{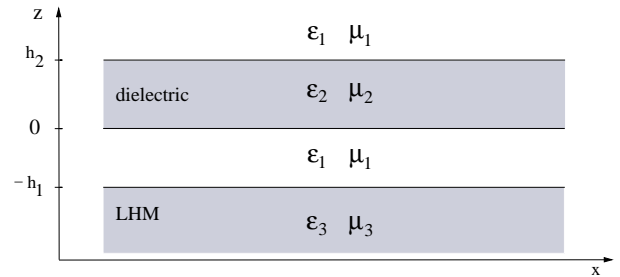}}
\caption{\label{fig:structure}Dielectric slab waveguide and the LHM interface separated by a distance $h_1$ and surrounded by a medium characterized by $\varepsilon_1$ and $\mu_1$.}
\end{figure}

Let us obtain the coupling with a symmetrical dielectric slab waveguide as depicted Fig. \ref{fig:structure}.
According to its dispersion relation, the thickness of the waveguide supporting a forward guided mode for the propagation constant $\alpha_0$, in TM polarization, is given by
\begin{equation}
h_2=\frac{2}{\gamma_2}\arctan\left(\frac{\varepsilon_2\kappa_1}{\varepsilon_1\gamma_2}\right),
\end{equation}
where $\kappa_1=\sqrt{\alpha_0^2-\varepsilon_1\,\mu_1\,k_0^2}$ and $\gamma_2=\sqrt{\varepsilon_2\,\mu_2\,k_0^2-\alpha_0^2}$.
For $\varepsilon_1=1$, $\mu_1=1$, $\varepsilon_2=3$ and $\mu_2=1$, the value of thickness $h_2$ is $0.2854\lambda$. In the following, we retain the above characteristics.

\begin{figure}[htb]
\centerline{\includegraphics[width=8cm]{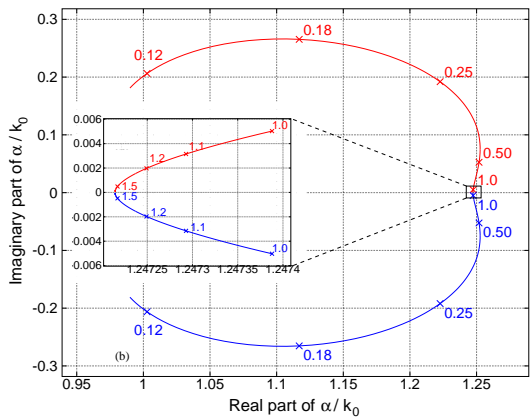}}
\caption{\label{fig:complexplane}Solutions of the dispersion relation in the $\alpha$ complex plane, for different values of $h_1$ in wavelength, the distance between the dielectric waveguide and the LHM surface. Inset, zoom in the region around $\alpha_0$.}
\end{figure}

\begin{figure}[htb]
\centerline{\includegraphics[width=8cm]{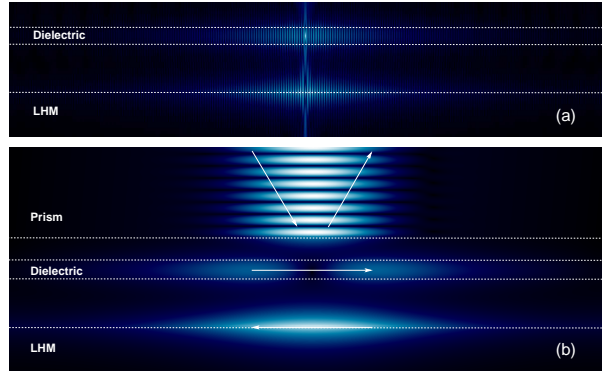}}
\caption{\label{fig:beam} Modulus of the field represented in a domain $100\,\lambda$ large, $5\,\lambda$ high. The distance between the dielectric slab and the LHM interface is $h_1=0.8\lambda$. (a) The punctual source, placed in the dielectric waveguide, excites two contrapropagative light wheels. (b) The light wheel is excited by a an incident gaussian beam (angle: $33.9^{\circ}$, waist: $10\,\lambda$ ) in a prism ($\varepsilon=5$, $\mu=1$). White arrows indicate the propagation direction of light. These images are obtained using the numerical method described in \cite{jeos}. }
\end{figure}

The above dielectric slab and LHM interface are brought close together, forming the structure presented in Fig. \ref{fig:structure}.
We have calculated the analytical expression of the dispersion relation of the whole structure, in TM polarization:
\begin{equation}
x_2\frac{\left[\exp(2j\gamma_2h_2)+X_2\right][\exp(2\kappa_1h_1)+X_3]}{[\exp(2j\gamma_2h_2)-X_2][\exp(2\kappa_1h_1)-X_3]}=-1,
\label{eq:dispersion}
\end{equation}
where $x_i=\kappa_i\varepsilon_1 / \kappa_1\varepsilon_i$ and $X_i=(1-x_i)/(1+x_i)$.
Figure \ref{fig:complexplane} shows, in the alpha complex plane, the solutions of the dispersion relation (Eq. (\ref{eq:dispersion})) for several values of distance $h_1$, the frequency being fixed.
When distance $h_1$ between the slab and the interface is large enough, the waveguides are independent, the dispersion relation is thus verified for $\alpha=\alpha_0$. But when distance $h_1$ decreases, they become coupled and two complex-conjugate solutions appear. For instance at $h_1=0.8\lambda$, $\alpha=(1.247994\pm0.012785i)k_0$.
These twin solutions have the same real part so that the corresponding modes cannot be excited separately by a source and they form a light wheel as shown in Fig. \ref{fig:beam}. 

The nonzero imaginary part of $\alpha$ means that the field $E(x,z,t)=E_0(z,t)\,\exp(i\alpha x)$ decays along the $x$ axis. Because the imaginary parts of the solutions are opposite, the twin modes decay in opposite directions. The size of the light wheel is moreover controled by $\Im(\alpha)$, a characteristic width of the light wheel being given by $L=\frac{2}{\Im(\alpha)}$.
Figure \ref{fig:complexplane}(a) shows that the smaller the distance $h_1$ is, the higher the imaginary part of the propagation constant is and so the smaller the light wheel will be.
However, Fig. \ref{fig:complexplane}(b) shows that the imaginary part of $\alpha$ presents a maximum reached for strong coupling for $h_1=0.17\lambda$. The light wheel can, thus, confine the light in a region as small as a characteristic width $L=1.17\lambda$.
Figure \ref{fig:beam}(a) shows the electromagnetic field created in the structure by a punctual source inside the dielectric layer. The interference pattern is due to the two contrarotative light wheels, which are excited and interfere. In Fig. \ref{fig:beam}(b), the light wheel is excited by a beam using evanescent coupling: a prism is placed above the dielectric slab and the structure is illuminated by a gaussian beam coming from above. Here, the field distribution is less intuitive. In particular, a dark zone appears just below the incident beam, in the center of the dielectric waveguide but not at the plasmonic interface. The presence of a similar dark zone has already been noticed but not explained \cite{tichit}.
We have applied the coupled-mode theory (CMT) \cite{pierce, yariv, huang} to this contradirectional coupling to get an analytical model able to account for this phenomenon.

Let us consider two independent guided modes whose complex amplitudes are $A$ and $B$.
Mode $a$ corresponds to a right traveling wave in the dielectric slab, whereas mode $b$ travels to the left on the LHM interface.
If the two guides are brought close together, they become coupled: energy is exchanged between them. 
Hence in this contradirectional coupling case and under phase matching conditions, the complex amplitudes $A$ and $B$ depend on $x$, obeying relations of the type \cite{yariv}
\begin{equation}
	\frac{dA}{dx} =\kappa^* B,
\label{eq:coupled1}
\end{equation}
\begin{equation}
	\frac{dB}{dx} =\kappa A,
\label{eq:coupled2}
\end{equation}
where $\kappa$ is the coupling coefficient. Its value is given by the imaginary part of the solution $\alpha$ of dispersion relation Eq. (\ref{eq:dispersion}). For weak coupling as well as for strong coupling, we indeed have $\kappa=i\Im(\alpha)$.

Here, the evanescent coupling used to excite mode $a$ is assumed to be weak so that the guided modes are unmodified by the presence of the prism.
An approach suggested by \cite{ulrich} is to consider the guided mode excited by a set of punctual sources situated inside the waveguide with an amplitude distribution along the $x$ direction given by the incident beam.
Each punctual source has an amplitude proportional to the amplitude of the incident field at the prism interface just above.

Let us then determine $A_1(x)$ and $B_1(x)$, the modes amplitudes created by a punctual source localized at $x=0$ whose the amplitude is equal to one, i.e., a source given by the expression $S_1(x)=\delta(x)$, where $\delta$ is the Dirac distribution. $A_1$ and $B_1$ can be seen as Green's functions \cite{ulrich}.
The solutions of Eqs. (\ref{eq:coupled1}) and (\ref{eq:coupled2}) are
\begin{equation}
	A_1(x) =H(x)\exp(-|\kappa|x) - H(-x)\exp(|\kappa|x),
\end{equation}
\begin{equation}
	B_1(x) =\exp\left(-i\frac{\pi}{2}\right)\left[H(x)\exp(-|\kappa|x) + H(-x)\exp(|\kappa|x) \right],
\label{eq:green}
\end{equation}
where $H$ is the Heaviside function. Notice the $\pi/2$ phase shift between the amplitudes $A_1$ and $B_1$ in each domains $x>0$ and $x<0$.

When the structure is excited by a source amplitude distribution $S(x)$, the amplitudes of the modes are obtained by convolution of the source expression $S$ with the Green's functions $A_1$ and $B_1$.
Thus, for a gaussian beam whose the waist size is $w$, i.e., an amplitude distribution $S(x)=S_0\,\exp\left(-x^2/w^2\right)$, we get
\begin{eqnarray}
A(x)&=&S_0\frac{\sqrt{\pi}}{2}\exp\left(\frac{\kappa^2w^2}{4}\right) \nonumber \\
& & \times \left[\exp(-|\kappa| x)\,\mathrm{erfc}\left(-\frac{x}{w}+\frac{|\kappa|w}{2}\right) \right. \nonumber \\
& & \left.
-\exp(|\kappa| x)\,\mathrm{erfc}\left(\frac{x}{w}+\frac{|\kappa|w}{2}\right)\right],
\label{eq:final}
\end{eqnarray}
\begin{eqnarray}
B(x)&=& S_0\frac{\sqrt{\pi}}{2}\exp\left(\frac{\kappa^2w^2}{4}-i\frac{\pi}{2}\right) \nonumber \\
& & \times \left[-\exp(-|\kappa| x)\,\mathrm{erfc}\left(-\frac{x}{w}+\frac{|\kappa|w}{2}\right) \right. \nonumber \\
& & \left. +\exp(\kappa x)\,\mathrm{erfc}\left(\frac{x}{w}+\frac{|\kappa|w}{2}\right) \right],
\label{eq:final2}
\end{eqnarray}
where erfc is the complementary error function.

\begin{figure}[htb]
\centerline{\includegraphics[width=7.8cm]{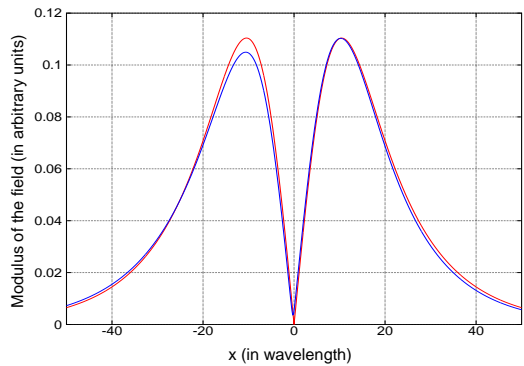}}
\caption{ Modulus of the field in the middle of the dielectric waveguide (blue curve) and $|A(x)|$ the modulus of the theoretical amplitude of the mode $a$ (red curve). Parameters of the structure and the beam are as in Fig. \ref{fig:beam}. $S_0$ is chosen arbitrarily.}
\label{fig:profiles}
\end{figure}

Figure \ref{fig:profiles} shows the modulus of the field distribution in the middle of the dielectric slab of Fig. \ref{fig:beam}(b). It is compared to the theoretical mode profile $|A(x)|$ computed for $\kappa=0.080330i$, given by the dispersion relation. Because the coupling with the prism is weak, the coupled modes are relatively undisturbed and Eq. (\ref{eq:final}) accurately describes the field in the slab. Both curves show a minimum for $x=0$ which corresponds to the dark zone in the middle of the dielectric waveguide.
The light wheel phenomenon can be summarized from Eqs. (\ref{eq:final}) and (\ref{eq:final2}) as follows: the guided mode is excited in the dielectric waveguide towards the right. It is then transferred by contradirectional coupling to the plasmonic backward mode with a $-\pi/2$ phase shift. There is no reason why the backward mode should undergo any phase change in $x=0$, and the energy is
transferred to the dielectric waveguide for $x<0$, with another phase shift of 
$-\pi/2$. In the dielectric slab, the right part and the left part of the
light wheel are thus in phase opposition. This cannot be seen in Fig. \ref{fig:beam}(a) because the source is ponctual. In Fig. \ref{fig:beam}(b), because the evanescent coupling is equivalent to a spatially extended source, the parts of the light wheel which are in phase opposition "overlap" in the dielectric slab and  a dark zone appears.

When we consider a lossy LHM, which is more likely, the light wheel and the dark zone still exist as shown in \cite{tichit}. In this case, the model can be simply extended by changing Eq. (\ref{eq:coupled2}) into $dB/dx =\kappa A - \kappa_l B$, where $\kappa_l$ is the extinction coefficient.

Finally, the complex plane analysis and the CMT associated with Ulrich's approach of evanescent coupling allow a very accurate description and a deep understanding of the light wheel phenomenon and its universal features. Beyond the fact that a light wheel can be used to confine light, this phenomenon can be used for beam reshaping \cite{jeos}, and an analytical model is in this context particularly
useful.


\end{document}